\documentclass{iau} 
\usepackage{graphicx}

\title[Extended HI disks in nearby spiral galaxies] %% give here short title %%
{Extended HI disks in nearby spiral galaxies}

\author[Albert Bosma]   %% give here short author list %%
{Albert Bosma$^1$
 }

\affiliation{$^1$Aix Marseille Universit\'e, CNRS, LAM 
(Laboratoire d'Astrophysique de Marseille) UMR 7326, 13388, 
Marseille, France \\ email: {\tt bosma@lam.fr} }

\pubyear{2016}
\volume{321}  %% insert here IAU Symposium No.
\jname{Formation and evolution of galaxy outskirts}
\editors{A. Gil de Paz, J. C. Lee \& J. H. Knapen, eds.}
\begin{document}

\maketitle

\begin{abstract}
In this short write-up, I will concentrate on a few topics of interest.
In the 1970s I found very extended HI disks in galaxies such as NGC 5055 and NGC 2841, out to 2 -  2.5 times the Holmberg radius. Since these galaxies are warped, a ``tilted ring model" allows rotation curves to be derived, and evidence for dark matter to be found. The evaluation of the amount of dark matter is hampered by a disk-halo degeneracy, which  can possibly be broken by observations of velocity dispersions in both the MgI region and the CaII region. 
\keywords{extended HI disks, warp, dark matter}
\end{abstract}

\firstsection % if your document starts with a section,
\section{Early HI work, warped HI disks, and the dark matter problem}

Several facts were discussed in the late 1950s concerning the HI emission in the outskirts of the Milky
Way: the HI disk is warped, and also flares as discussed in \cite[Gum et al. (1960)]{Gum60}. 
Since M31 approaches the Milky Way, 
\cite[Kahn \& Woltjer (1959)]{KW59} argued that dark matter has to be present in the Local Group. 
They thought it to be intergalactic gas, through which the Milky Way moves, thus causing the warp.
For M31, a study by \cite[Van de Hulst, Raimond \& Van Woerden (1957)]{Hul57} hinted at high 
mass-to-light ratio material in the outer parts, but the data were as yet uncertain.
In the 1970s, dark matter around spiral galaxies became a subject in itself, thanks in part
to HI studies in other galaxies. This has been reviewed often, so here I only 
mention some early data on large HI outskirts.
%, which established the warp problem, and the very extended HI disks around some galaxies.

\cite[Rogstad, Lockhart \& Wright (1974)]{Rog74} found a large outer HI envelope in M83, and
modelled it with a tilted ring model where the outer rings were not in the same plane as the inner disk. A direct
warp was found by \cite[Sancisi (1976)]{San76} in the edge-on galaxies NGC 5907 and NGC 4565.  In my
thesis work (\cite[Bosma 1978]{Bos78}), I found several galaxies for which the HI 
extends much farther out than the optical image, the best case being the galaxy NGC 2841, cf. Figure 1.
The warp was modelled with tilted rings, now using the circular velocity in each ring as a free
parameter. Since NGC 2841 does not have an obvious companion, it was concluded that the warp was not 
due to tidal interaction, as for NGC 5907 and NGC 4565.
For NGC 5055, the extended HI disk was also found to be warped, but
on a deep IIIaJ image there is a irregular patch of optical
emission in the galaxy outskirts (\cite[Bosma 1978]{Bos78}, Fig. 4.5.1). This galaxy became much later a prototype 
extended UV disk of type 1 (\cite[Thilker et al. 2007]{Thil07}), and 
CO emission in the patch has been reported by \cite[Dessauges-Zavadsky et al. (2014)]{Des14}.  These data,
combined with the images in \cite[Gonzalez-Delgado et al. (2010)]{Gon10}, make capture of a dwarf galaxy a
likely scenario.
Finally, I collected rotation curves of other galaxies from new
Westerbork data, either by myself or by Sancisi, or from data already in the literature. With P.C. van der Kruit, we collected and analysed deep IIIaJ images obtained with the 48" Palomar Schmidt, and from mass modelling we 
found that the local mass-to-light ratios increase to more than 200 in several cases (\cite[Bosma 1978]{Bos78}, Figs. 6.1 and 7.6, and \cite[Bosma \& van der Kruit 1979]{BK79}).

\begin{figure*}[ht]
\centering
\includegraphics[scale=0.203, angle=0.0]{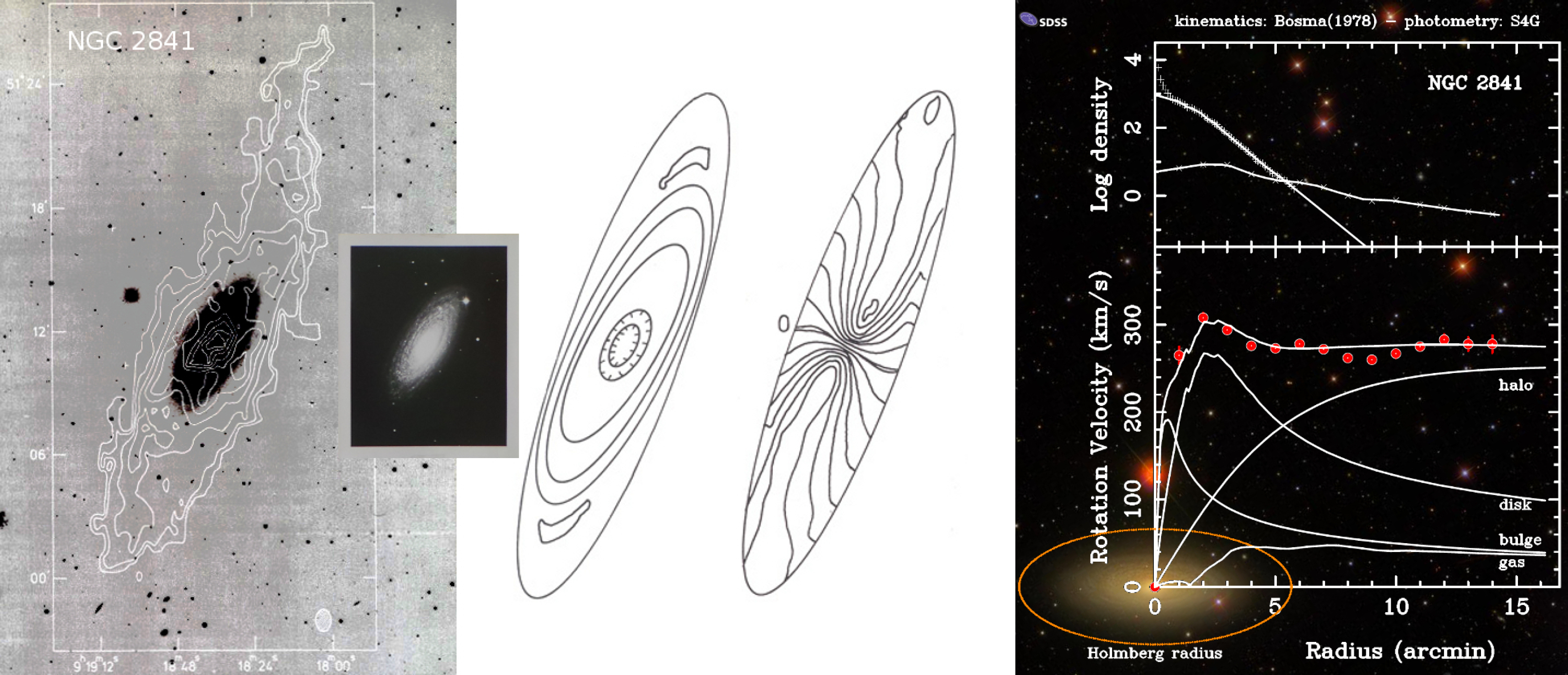}
\caption{At left, the HI distribution in the galaxy NGC 2841 observed by Bosma (1978), overlaid on 
an deep IIIaJ image kindly provided by H.C. Arp; the inset shows an image similar to the one 
in Sandage's (1961) Hubble Atlas. The middle panels show the warp model.  At right a mass model
adjusted to the rotation curve in Bosma (1978), calculated as described in Athanassoula et al. (1987), using surface photometry from the
S4G survey (Munoz-Mateos et al. 2015), overlaid on scale on an image obtained from the SDSS. The ellipse centered 
on the galaxy image outlines the Holmberg (1958) dimensions of the galaxy. 
}
\label{fig:n2841}
\end{figure*}

\section{On the disk-halo degeneracy}

\begin{figure*}[b]
\begin{center}
\includegraphics[scale=0.30, angle=0.0]{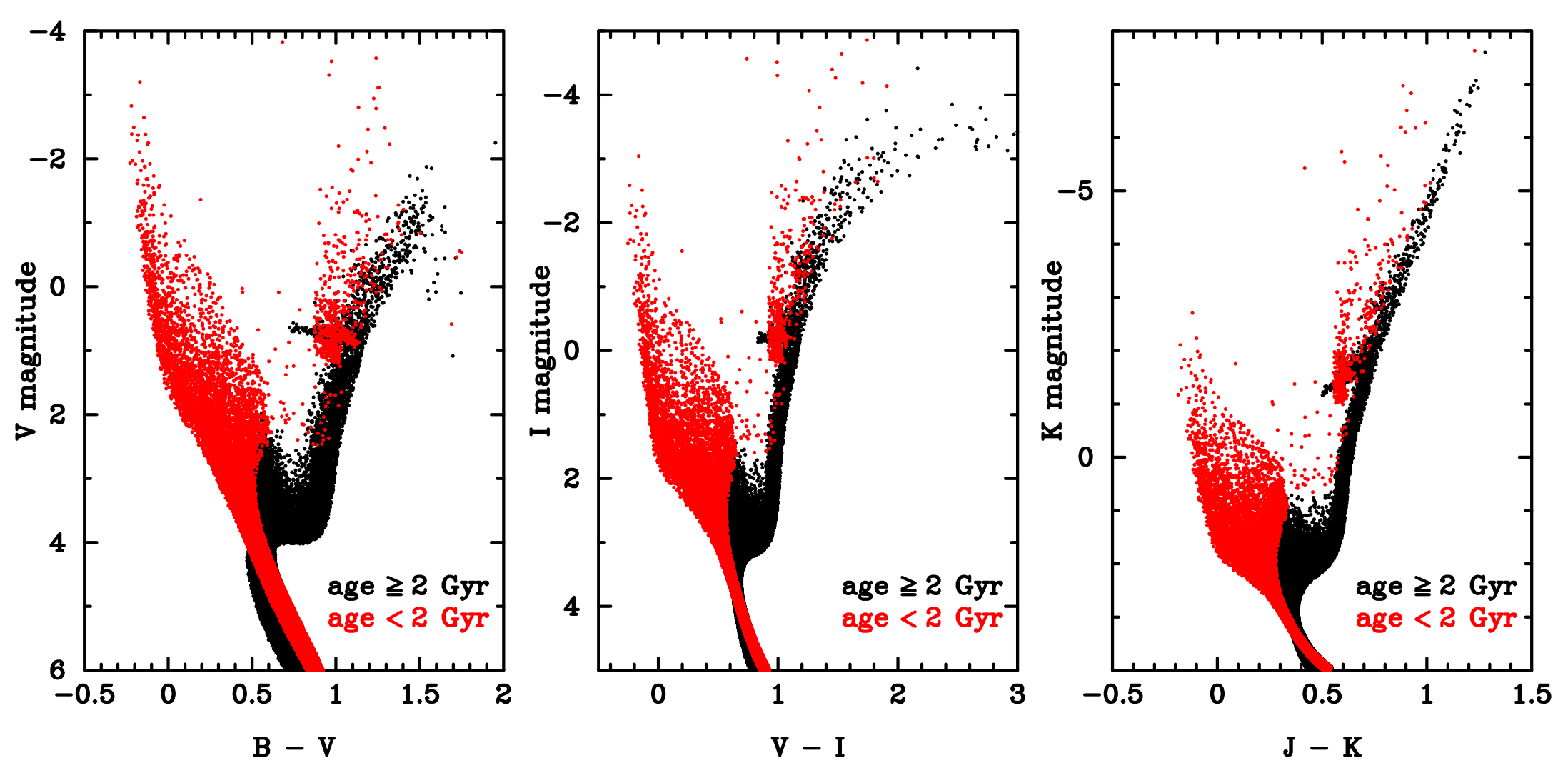}
\end{center}
\caption{CMDs calculated with IAC-STAR (\cite[Aparicio \& Gallart 2004]{AG04}) in several wavelength regions.
Further into the infrared the older stars are dominating the light.~~~~~~~~~~~~~~~~~~~~~~~~~~~~~~~}
\label{fig:veldisp}
\end{figure*}

The question then shifted from ``is dark matter present?" to "how much 
of it is there, and what is its distribution?".
This led to the formulation of the disk-halo degeneracy problem (\cite[Van Albada \& Sancisi 1986]{vAS86}), which
states that there is no tight constraint on the mass-to-light ratio of the disk (now taken constant as function of radius).
Further constraints come from dynamical arguments, e.g. the 
ability of a galactic disk to amplify two-armed spiral structure with the swing amplifier mechanism
(\cite[Athanassoula, Bosma \& Papaioannou 1987]{ABP87}), leading to a range of
M/L values (in Figure 1 a maximum disk is used). Other constraints have been used, such as spiral structure models (\cite[Kranz, Slyz \& Rix
2003]{KSR03}), and the strength of shocks in the gas flow in barred spirals 
(\cite[Weiner, Sellwood \& Williams 2001]{Wei01}). 
Most studies agree that for large, high surface brightness
galaxies the disk is near maximum, for smaller, lower surface brightness galaxies the disk becomes sub maximum,
and for dwarf disk galaxies the dark matter dominates in the inner parts.

Studies using velocity dispersion data, however, find that even bright galaxies have
sub maximum disks. This conclusion of the DiskMass project (\cite[Bershady et al. 2011]{Ber11}) 
was already indicated by the work of \cite[Bottema (1993)]{Bot93}. Velocity dispersions could be
biased, however,  by the presence of low-velocity dispersion high luminosity young 
stars, as discussed in \cite[Bosma (1999)]{B99}.
This has been further quantified recently by \cite[Anyian et al. (2016)]{Any16}, who use colour-magnitude diagrams to strengthen the argument. In Figure 2 I take their argument further: I plot CMDs in the blue (B vs. B-V), near infrared
(I vs. V-I) and further in the infrared (K vs. J-K), to show that it can be expected that, if the young stellar populations
really have a significant effect on the velocity dispersions, it matters in which wavelength region 
the data are taken: (slightly) higher velocity dispersions should be found 
for data taken in the CaII region ($\lambda$ $\sim$ 8500 \AA) compared to the MgI region ($\lambda$ $\sim$ 5175 \AA). This can be tested with current
and future data using Integral Field Units with sufficient spectral resolution.

\acknowledgements{This work was supported by the DAGAL network from the People Programme (Marie
Curie Actions) of the European Union's Seventh Framework Programme FP7/2007-2013/
under REA grant agreement PITN-GA-2011-289313, and by the CNES (Centre National
d'Etudes Spatiales - France). This work has
made use of the IAC-STAR Synthetic CMD computation code. IAC-STAR is supported and maintained by the computer division of the Instituto de Astrof\'isica de Canarias.}

\end{document}